\documentclass[%
 aip,
 amsmath,amssymb,
 reprint,%
]{revtex4-2}

\usepackage{graphicx}
\usepackage{dcolumn}
\usepackage{bm}

\usepackage[utf8]{inputenc}
\usepackage[T1]{fontenc}
\usepackage{mathptmx}
\usepackage{etoolbox}

\makeatletter
\def\@email#1#2{%
 \endgroup
 \patchcmd{\titleblock@produce}
  {\frontmatter@RRAPformat}
  {\frontmatter@RRAPformat{\produce@RRAP{*#1\href{mailto:#2}{#2}}}\frontmatter@RRAPformat}
  {}{}
}%
\makeatother

\usepackage{amsfonts}
\usepackage{subcaption}
\usepackage{url,hyperref}
\usepackage{cleveref}
\usepackage{soul}
\usepackage[english]{babel}	

\newcommand{\SI}[2]{\ensuremath{#1\,\mathrm{#2}}}
\newcommand{\SIU}[1]{\ensuremath{\mathrm{#1}}}

\newcommand{\Vector}[3]{\begin{pmatrix}#1\\#2 
\ifthenelse{\equal{#3}{}}{}{\\#3}\end{pmatrix}}

\newcommand*\euler{\mathrm{e}}

\begin{document}

\preprint{AIP/August2022}

\title{Guided Electromagnetic Discharge Pulses Driven by Short Intense Laser Pulses: Characterisation and Modelling}

\author{M.~Ehret}
\email[M. Ehret currently at CLPU, Villamayor, Spain, ]{mehret@clpu.es}
\affiliation{Universit\'{e} de Bordeaux, CNRS, CEA, CELIA (Centre Lasers Intenses et Applications),
UMR 5107, Talence, France}
\affiliation{Institut f\"{u}r Kernphysik, Technische Universit\"{a}t Darmstadt, Darmstadt, Germany}

\author{M.~Bailly-Grandvaux}
\affiliation{Universit\'{e} de Bordeaux, CNRS, CEA, CELIA (Centre Lasers Intenses et Applications),
UMR 5107, Talence, France}

\author{Ph.~Korneev}
\affiliation{National Research Nuclear University MEPhI, Moscow, Russian Federation}
\affiliation{P.N. Lebedev Physics Institute, Russian Academy of Sciences, Moscow, Russian Federation}

\author{J.J.~Api\~{n}aniz}
\affiliation{CLPU (Centro de L\'{a}seres Pulsados), Villamayor, Spain}

\author{C.~Brabetz}
\affiliation{Plasma Physik/PHELIX, GSI Helmholtzzentrum f\"{u}r Schwerionenforschung GmbH, Darmstadt,
Germany}

\author{A.~Morace}
\affiliation{Institute of Laser Engineering (ILE), Osaka University, Osaka, Japan}

\author{P.~Bradford}
\affiliation{Department of Physics, York Plasma Institute, University of York, Heslington, UK}

\author{E.~d'Humi\`{e}res}
\affiliation{Universit\'{e} de Bordeaux, CNRS, CEA, CELIA (Centre Lasers Intenses et Applications),
UMR 5107, Talence, France}

\author{G.~Schaumann}
\affiliation{Institut f\"{u}r Kernphysik, Technische Universit\"{a}t Darmstadt, Darmstadt, Germany}

\author{V.~Bagnoud}
\affiliation{Plasma Physik/PHELIX, GSI Helmholtzzentrum f\"{u}r Schwerionenforschung GmbH, Darmstadt,
Germany}

\author{S.~Malko}
\affiliation{CLPU (Centro de L\'{a}seres Pulsados), Villamayor, Spain}

\author{C.~Matveevskii}
\affiliation{National Research Nuclear University MEPhI, Moscow, Russian Federation}

\author{M.~Roth}
\affiliation{Institut f\"{u}r Kernphysik, Technische Universit\"{a}t Darmstadt, Darmstadt, Germany}

\author{L.~Volpe}
\affiliation{CLPU (Centro de L\'{a}seres Pulsados), Villamayor, Spain}

\author{N.~C.~Woolsey}
\affiliation{Department of Physics, York Plasma Institute, University of York, Heslington, UK}

\author{J.J.~Santos}
\affiliation{Universit\'{e} de Bordeaux, CNRS, CEA, CELIA (Centre Lasers Intenses et Applications),
UMR 5107, Talence, France}


\date{\today}

\begin{abstract}
Strong electromagnetic pulses (EMP) are generated from intense laser interactions with solid-density targets, and can be guided by the target geometry, specifically through conductive connections to the ground. We present an experimental characterization, by time- and spatial-resolved proton deflectometry, of guided electromagnetic discharge pulses along wires including a coil, driven by \SI{0.5}{ps}, \SI{50}{J}, \SI{10^{19}}{W/cm^2} laser pulses.
Proton-deflectometry data allows to time-resolve first the EMP due to the laser-driven target charging and then the return EMP from the ground through the conductive target stalk. Both EMPs have a typical duration of tens of \SIU{ps} and correspond to currents in the \SIU{kA}-range with electric-field amplitudes of multiple \SIU{GV/m}.
The sub-\SIU{mm} coil in the target rod creates lensing effects on probing protons, due to both magnetic- and electric-field contributions. This way, protons of \SI{10}{MeV}-energy range are focused over \SIU{cm}-scale distances. 
Experimental results are supported by analytical modelling and high-resolution numerical particle-in-cell simulations, unraveling the likely presence of a surface plasma, which parameters define the discharge pulse dispersion in the non-linear propagation regime. 

\end{abstract}


\maketitle

\section{Introduction}

Electromagnetic pulse (EMP) emission resulting from pulsed laser interaction with solid targets is reported for a large range of laser parameters with energies from \SI{10}{mJ} to \SI{1}{MJ} and intensities from \SI{10^{14}}{W/cm^2} to \SI{10^{20}}{W/cm^2} in the relativistic regime \cite{Qu:2009,To:2015,Ah:2016,Ka:2016,Eh:2017,Consoli:2020}.
In parallel, the guiding of such EMPs as an electrodynamic lensing technique is being pursued by groups interested in focusing and post-acceleration of laser-accelerated particle beams \cite{Ka2016-2,Ba2020}. Though these first particle-beam lensing experiments have considerably advanced our knowledge, the physical mechanisms responsible for the formation and propagation of guided EMP are not entirely understood. We present here experimental evidence of EMP bound to the target geometry and propose a model based on target discharge and geometry able to predict the EMP peak amplitude and dispersion relation. Furthermore, we follow the return current dynamics after the discharge pulse.\par

During laser interaction with a solid foil target a positive potential builds up close to the irradiated surface. This is due to laser-accelerated fast electrons that overcome the potential barrier and escape \cite{Po:2015-2}. Electron charge extraction with ultra-intense (\SI{10^{18}}{W/cm^2} -- \SI{10^{20}}{W/cm^2}) sub-ps laser pulses ensues from mechanisms such Brunel-type resonance absorption \cite{Br:1987,Wi:1997} and ponderomotive $j \times B$ acceleration \cite{Wi:1992,Pu:1998}. This gives rise to the generation of intense broadband EMPs propagating into the space surrounding the target, spectrally ranging from radio frequencies \cite{Pe:1977} to X-rays \cite{Co:2009}.\par

Another fraction of electrons is accelerated forward into the target {bulk}. The most relativistic can leave the target at the rear side yielding a supplementary positive potential {\cite{Ga2020}}. Potential created after both front- and rear side electron escape {spreads} along the target surface \cite{Ha:2000}, where according to particle-in-cell (PIC) simulations, a net non-zero charge density forms only within the skin depth \cite{Qu:2009}.\par

This charge density does not spread neither instantly nor uniformly over the whole target body. The PIC simulations reveal a discharge wave with time-scale of tens of \SIU{ps} traveling along the target. In experiments with sub-ps laser driven wire targets, discharge pulses were observed several \SIU{mm} away from the laser-interaction region \cite{To:2015}. These pulses show weak dispersion and attenuation during their linear propagation, but clear losses after reflection at the open end of a wire target. Modelling the pulse with a Sommerfeld wave \cite{To:2015,Brantov2018} reproduces qualitatively the observed strong radial electric- (E-) and azimuthal magnetic- (B-) field components. The long travel range with no considerable dispersion or attenuation was confirmed experimentally \cite{Ah:2016}.\par

The present investigation aims at the {experimental characterization and physical understanding of} the formation and evolution of such a discharge pulse and the subsequent return current dynamics. Particularly, we {develop} a new analytical model capable to describe and understand the observed propagation velocity and fine structure of the discharge pulses. The model predictions are consistent with PIC simulations, where we can discriminate EMP in free space from target-bound EMP, and check the assumptions made for the modelling. We find that the positive potential evolves and gives rise to a pulsed electric current within the skin depth of the target rod, propagating with a group velocity close to the speed of light and bearing E- and B-fields. By using coil-shaped rods, the fields can be explored as lensing platforms for laser-accelerated beams of charges. We will consider here a simple scheme based on flat targets laser-cut from a thin metallic foil.

{The paper is organized as follows, firstly experimental results are presented and a heuristic approach is applied to quantify the evolution of the target-bound discharge pulses. Secondly, PIC simulations are presented supporting the basic assumptions made for the heuristic analysis and allow further insights into the discharge pulse dynamics. Thirdly, an analytical approach to model the discharge pulse dispersion is presented, which agrees with the observed time-scales. The dynamics of pulsed return currents from the ground is explored in a fourth section. Finally, we conclude and comment on how EMP discharges can be used in future experimental applications.}

\section{Experiment}

\begin{figure}
	\includegraphics[width=0.9\columnwidth]{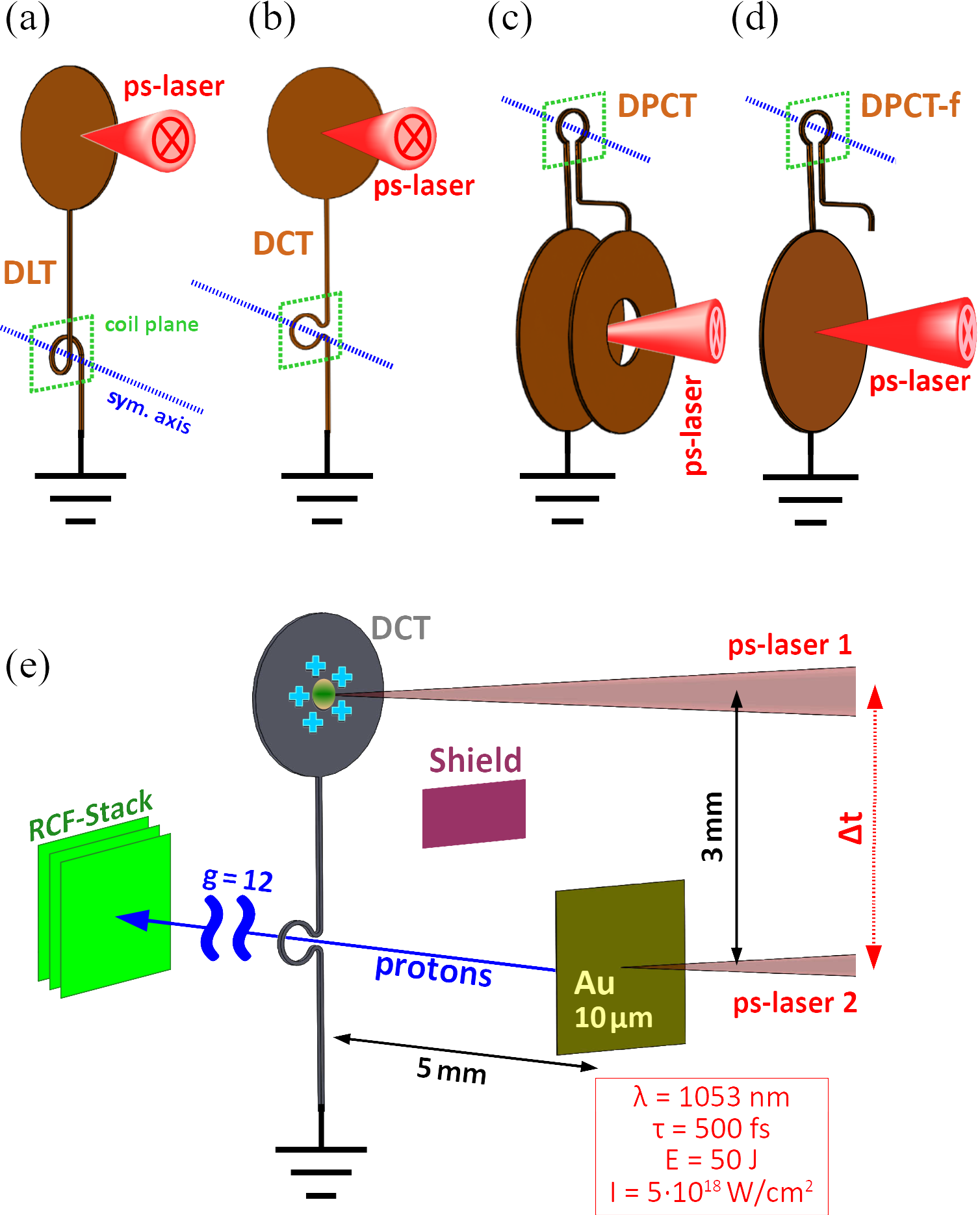}%
	\caption{Targets are laser-cut in one piece from \SI{50}{\mu m}-thick Cu foils. All types comprise an interaction disk and a loop of \SI{250}{\mu m}-radius in their \SI{50}{\mu m}-side squared grounding stalk: (a) Disk Loop Targets (DLT) with a helix-shaped loop, (b) Disk $\Omega$-Coil Targets (DCT), (c) Double-Plate Coil Targets (DPCT) with a $\Omega$-shaped loop and (d) Double-Plate Coil Targets without front plate (DPCT-f) with a $\Omega$-shaped loop. (e) The set-up for target discharge and proton-deflectometry probing relies on two identical \SIU{ps}-laser pulses with adjustable delay $\Delta t$. \SIU{ps}-laser 1 induces a target discharge on the interaction disk and \SIU{ps}-laser 2 drives a proton beam from the rear surface of a \SI{10}{\mu m}-thick Au foil. Deflectometry images are obtained from the proton dose deposition over a stack of RCF. A \SI{100}{\mu m}-thick Tantalum shield blocks the direct line of sight between both laser interaction regions.\label{fig:DCTprobing}}
\end{figure}

The experiment \cite{Eh:2017} was carried out at GSI with the Petawatt High Energy Laser for Heavy Ion Experiments (PHELIX) \cite{Ba:2016}. We report on target discharges driven by laser pulses of \SI{500}{fs} duration, \SI{50}{J} energy and intensities of \SI{5\cdot 10^{18}}{W/cm^2}. The targets are laser-cut in one piece from \SI{50}{\mu m} thick flat Cu foils. All types of the four different targets comprise an interaction disk and a wire connection to the ground of \SI{50^2}{\mu m^2} squared-section that includes a loop-feature of \SI{250}{\mu m}-radius, {as} depicted in \cref{fig:DCTprobing}: (a) Disk Loop Targets (DLT) with a helix-shaped loop, (b) Disk $\Omega$-Coil Targets (DCT) and (c) Double-Plate Coil Targets (DPCT) with a $\Omega$-shaped loop. DPCT have a more complex geometry with two parallel disks connected by the loop-shaped wire: the laser pulse passes through a hole in the front plate to drive the discharge from the rear plate. This geometry is simplified for a fourth target type: (d) DPCT-f has only the laser-irradiated plate, resulting in an open ended wire on one side of the loop. The interaction {disks} of DLT and DCT {have} a diameter of \SI{2}{mm}, DPCT and their derivation DPCT-f have \SI{3}{mm} diameter disks.\par

\begin{figure*}[t]
	\includegraphics[width=\textwidth]{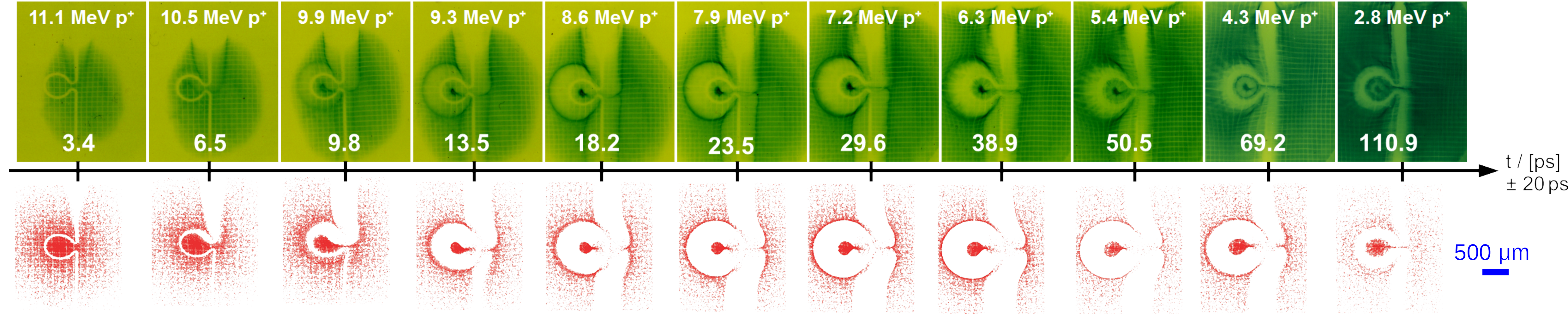}%
\caption{RCF proton imprints for different probing times within one shot using a DCT (top). Corresponding synthetic images (bottom) are obtained by coupled simulations of the dynamic target discharge pulse and test-particle probing. The laser pulse impacts in the target disk above the field of view. The discharge pulse is coming from the top guided by the target geometry. The un-driven $\Omega$-Coil (upper left RCF) has a diameter of \SI{500}{\mu m}. All images have the same spatial scale indicated with a scale bar projected to the coil position. The witness mesh is positioned between proton source and coil - its periodicity projects to \SI{106}{\mu m} in the coil plane. For late times features like a ring-shaped caustic inside the coil and filaments are visible.\label{fig:timeresolved}}
\end{figure*}

\begin{figure}
	\includegraphics[width=\columnwidth]{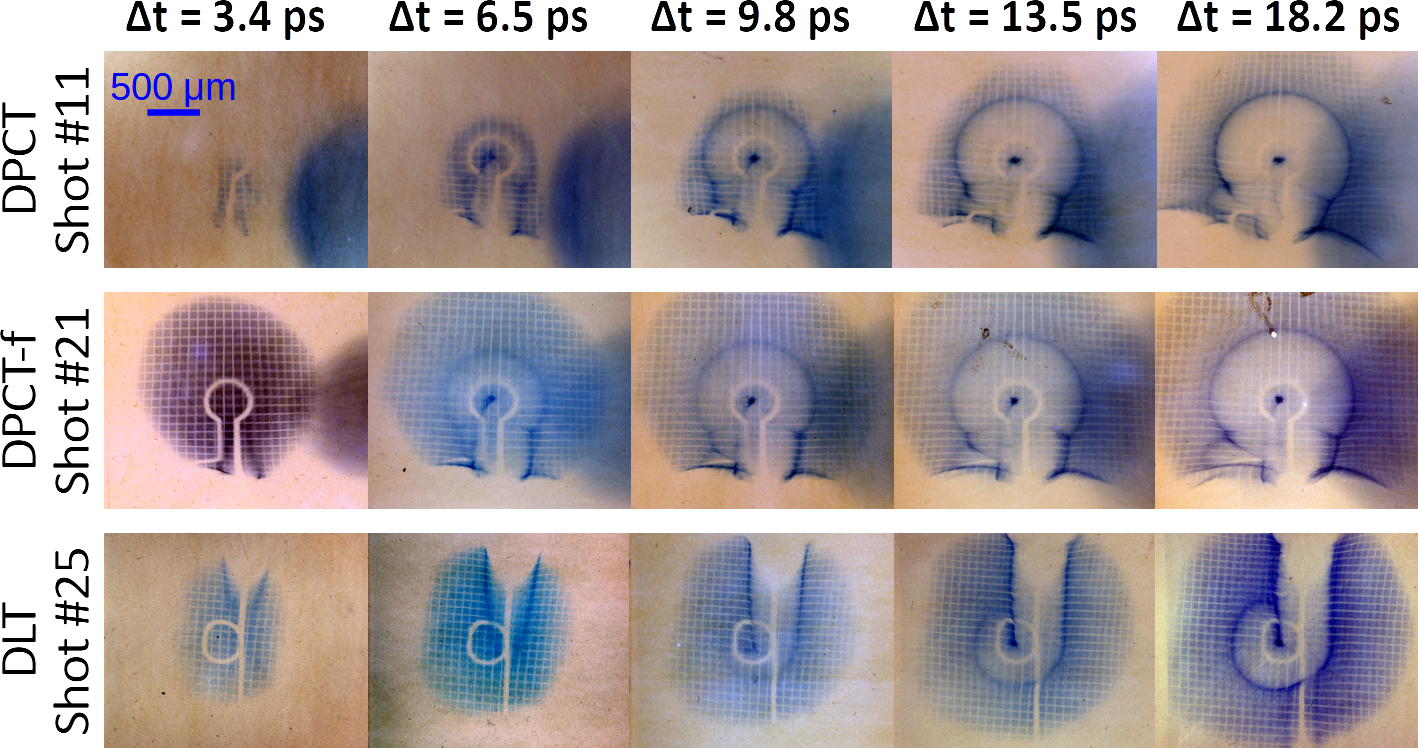}%
	\caption{Proton deflectographs at distinct probing times for three shots using respectively a DPCT (top), DPCT-f (mid) and DLT (bottom). The discharge pulse comes from the laser interaction region and is guided by the target geometry. The initial diameter of the $\Omega$-Coil is \SI{500}{\mu m} and all images have the same spatial scale, with the witness mesh having a periodicity of \SI{106}{\mu m} in the coil plane.\label{fig:velocities}}
\end{figure}

The discharge \SIU{ps} time- and \SIU{mm} spatial-scales were captured by proton-deflectometry. The probing proton particle beam is accelerated via Target Normal Sheath Acceleration (TNSA) using a second PHELIX laser beam portion similar to the discharge target driver, with an adjustable temporal delay. After crossing the target region of interest (ROI), the protons' deflections are imaged over a stack of Radiochromic Films (RCF), \cref{fig:DCTprobing} (e). Due to the characteristic Bragg peak of proton-energy absorption, each successive RCF image corresponds to the proton imprint of a small range of raising energy, therefore of different decreasing time-of-flight between the proton source and the ROI.\par

Deflectometry image data from a single shot are shown in \cref{fig:timeresolved} with a DCT. We obtained similar results with all target types, see \cref{fig:velocities}. The straight wire sections and the coil feature are clearly visible. For early times the target appears to be unaltered: deflections of protons then purely result from scattering in the solid density wire. Strong deflections away from the target rod with sharp caustics appear a few \SIU{ps} after target driving. After deflections reach their maximum, they decrease slowly back to zero, which conveys that they are not due to a thermally expanding target, but instead to a travelling EMP which wavefront propagation along the target wire is clearly {evidenced} through the image sequence.\par

For this very shot, given the chosen delay between the two laser pulses, \SI{8.6}{MeV} and \SI{9.3}{MeV} protons are those experiencing the peak amplitude of the propagating discharge pulse, as inferred from the larger horizontal deflections perpendicularly to the straight regions of the target rod. For both the corresponding RCF layers, one observes an enhancement of the proton beam dose deposition in proximity of the loop's symmetry axis, interpreted as the result of focusing trajectories after crossing the loop region. 

Though, in a non-stationary situation, when fields are changing rapidly for all positions of the proton path, the proton beam spectrum may change. The exact quantification of how much the emittance changes would require knowledge about the particle phase-space measured at three consecutive distances, which, for our setup, would only be achieved in three different shots assuming a perfect shot-to-shot reproducibility, to our best knowledge not yet feasible at high power laser facilities. To that, one would need to add identical reference shots without driving the coil targets. Despite these experimental limitations, we nevertheless propose a rough estimation of the transverse emittance perpendicular to the coil axis out of two distinct laser shots with the RCF stack placed at two different distances. We assume beam larminarity within narrow energy bins and beam focusing far behind the RCF stack. For these two shots the delay between the lasers was tuned to reach peak discharge amplitude at the probing time of \SI{6.3}{MeV} protons. We observed that the transverse emittance of those protons passing through the loop is reduced by a factor $\approx 3$ compared to reference shots without driving the coil: from initially $\SI{(1.59 \pm 0.05)}{mm \cdot mrad}$ to $\SI{(0.5 \pm 0.1)}{mm \cdot mrad}$. Note that changes of beam emittance can not arise with quasi-static electromagnetic lensing. Therefore, our results suggest longitudinal post-acceleration on timescales shorter than the transit time, changing the spectrum of the proton beam. These observations further highlight the non-stationary character of the guided discharges.

\subsection{Evolution of the Discharge Pulse}

\begin{table}
 \caption{Group velocity of the wave front along straight sections of the Cu wire in units of the speed of light $c$. Laser parameters and target material were not intentionally varied. Shot \#25 allows observation of the wave front in two consecutive pairs of RCF, yielding two measurements that agree within the margins of their uncertainties.\label{tab:phasespeed}}
 \begin{ruledtabular}
 \begin{tabular}{l|r|c|r}
 Shot \# & Discharge-Target 	& Target Type	& Group Velocity $v_\mathrm{g}$ \\
 		   & Driver Energy		&						&										\\
 11         & \SI{51.5}{J}		& DPCT				& $(0.77 \pm 0.10) \cdot c$ \\
 21         & \SI{51.5}{J}		& DPCT-f   			& $(0.82 \pm 0.05) \cdot c$ \\
 22         & \SI{47}{J}		& DCT    			& $(0.82 \pm 0.18) \cdot c$ \\
 37         & \SI{53.5}{J}		& DCT 				& $(0.78 \pm 0.20) \cdot c$ \\
 41         & \SI{41.4}{J}		& DCT				& $(0.95 \text{~}_{-0.10}^{+0.05}) \cdot c$ \\
 25         & \SI{44.3}{J}		& DLT 				& $(0.80 \pm 0.08) \cdot c$ \\
 25         &	\SI{44.3}{J}		& DLT				& $(0.78 \pm 0.09) \cdot c$ \\
 39         & \SI{51}{J}		& DLT   			& $(0.81 \text{~}_{-0.25}^{+0.19}) \cdot c$ \\
 \end{tabular}
 \end{ruledtabular}
\end{table}

{Seven shots allowed to see the wave front propagation imprinted on consecutive RCF. The measured mean group velocity of the wave front along the Cu-target rod is $(0.82 \pm 0.06) \cdot c$, with the minimum value $(0.77 \pm 0.10) \cdot c$ and the maximum value $(0.95 \text{~}_{-0.10}^{+0.05}) \cdot c$.} All measurements are given in \cref{tab:phasespeed}. The variation may be due to shot-to-shot differences in effective laser power and in target surface quality issuing from the laser cutting. The observed relativistic velocity close to that of light suggests the electromagnetic nature of the propagating wave, where electric and magnetic components are of a similar value.

\begin{figure}[hb]
	\includegraphics[width=\columnwidth]{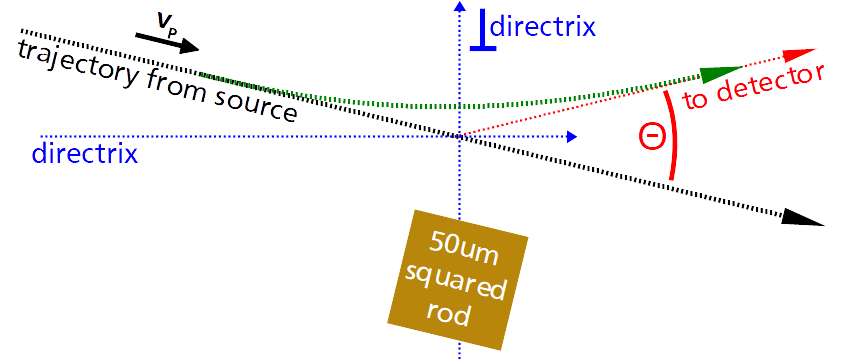}%
	\caption{The geometry of the probing proton deflection around the target rod.  \label{fig:geometryforderivation}}
\end{figure}

In order to reconstruct the discharge pulse amplitude versus time, we first deduce an approximation of a linear charge-density distribution ${\lambda (\vec{x}, t)}$ {yielding} the electric component of the discharge pulse. We assume that the {E-field} has a stronger influence than the B-field on radial (horizontal) deflections of probing protons along the straight target rod, for the accessed TNSA proton energy range of \SI{1 - 20}{MeV} (well below relativistic values). The proton deflection angle $\Theta (\vec{x},t )$, determined from the caustics imprinted on each RCF layer, is at maximum $\approx 3^{\circ}$. The deflection angle results of an radial acceleration with respect to the rod, as sketched in \cref{fig:geometryforderivation}. As first approximation, we neglect changes in the velocity component parallel to the directrix. Secondly, we set equal the norm of in- and outgoing velocity vector for trajectories from negative and positive infinity. In the limit of small deflection angles, we derive for the charge-density distribution: 

\begin{equation}\label{eq:lambda} 
\lambda \approx \frac{4 \pi \epsilon_0}{q_\mathrm{p}} \cdot \frac{m_\mathrm{p} v_\mathrm{p}^2}{2} \cdot \frac{\Theta}{\pi}
\end{equation}

\noindent with $m_\mathrm{p}$, $q_\mathrm{p}$ and $v_\mathrm{p}$ being the probing proton mass, charge and velocity respectively, and $\epsilon_0$ the vacuum permitivity. With a deflection of $3^{\circ}$ for protons of \SI{10}{MeV}, we obtain {$\lambda \approx $ \SI{20}{nC/mm}}.

\begin{figure}[htb]
	\includegraphics[width=\columnwidth]{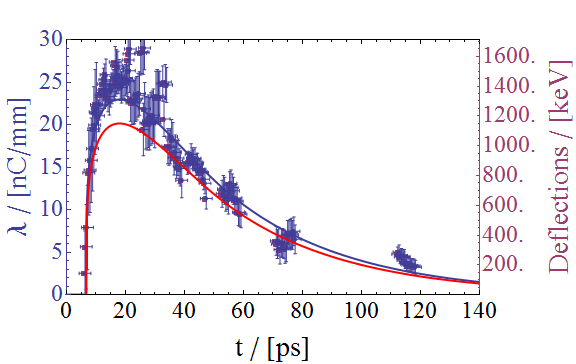}%
	\caption{Evolution of the discharge pulse. Blue symbols yield values extracted upon electrostatic assumptions from deflectometry data in \cref{fig:timeresolved} using \cref{eq:lambda} around straight wire sections of a DCT. $t=\SI{0}{ps}$ corresponds to the driver laser impact on the target disk. For reference, proton probing times and measurement positions are converted to the respective arrival time of the discharge pulse at the $\Omega$-Coil, assuming constant group velocity. The time uncertainty comprises the range of proton energies imprinted within one active RCF layer and the time the probing protons are in vicinity of the deflecting charge distribution, which is assumed to be of the order of the void around the wire. The gain of proton kinetic energy perpendicular to the probing axis is indicated on the plot's right hand side ordinate axis. A fit using \cref{eq:fit} (blue curve) yields an integrated charge of $Q_0=$ \SI{(342 \pm 5)}{nC}. Upon electrodynamic assumptions, data fits best to synthetic deflectograms when the former fit is renormalized to $Q_0=$ \SI{300}{nC} (red curve). \label{fig:peak}}
\end{figure}

Accordingly, blue symbols in \cref{fig:peak} relate the temporal evolution of the radial deflections (quantified in units of raise in the proton energy component perpendicular to the probing axis; right-hand-side ordinates) with the deduced evolution of the charge distribution (left-hand-side ordinates). {The temporal axis is given with respect to target drive, and individual measurements are compared using the respective proton arrival time at the $\Omega$-coil}. The peak of \SI{\approx 25}{nC/mm} is reached at \SI{10}{ps} and has a FWHM of \SI{\approx 50}{ps}. We observe an exponential decay after the peak. We found a modified continuously differentiable Weibull function with purely exponential tail to fit the data, reading

\begin{equation}\label{eq:fit} 
f(x) = \frac{Q_0}{k + \euler - 1} \cdot 
	\begin{cases} 
		0 &\mbox{if }  x \le 0 \\
		\frac{k}{\sigma} \left( \frac{x}{\sigma} \right)^{k-1} \exp \left[ 1 - \left( \frac{x}{\sigma} \right)^k \right] & \mbox{if } x \le \sigma \\
		\frac{k}{\sigma} \exp \left[ 1 - \left( \frac{x}{\sigma} \right) \right] & \mbox{if } x > \sigma \\
	\end{cases} 
,
\end{equation}

\noindent where $Q_0$ denotes the normalization factor of the function representing the total target discharge and $\mathrm e$ is {Euler's number}. The parameters fit to ${k= (1.25 \pm 0.02)}$ and ${\sigma= \SI{(10.1 \pm 0.3)}{mm}}$. Integration of the fitting blue curve corresponds to an equivalent total target discharge of \SI{Q_0=(342 \pm 5)}{nC}.

\subsection{Dynamic Analysis}

In order to reproduce synthetically the experimental RCF results, we performed dynamic test-particle transport and electromagnetic-field simulations with the Particle Field Interaction (PAFIN) code \cite{Eh:2015}. The code processes any combination of magnetic- and electric fields with the possibility of defining current density and charge density distributions, and it allows implementation of analytical solutions for fields as time-varying functions. After generation of a particle beam with a given phase space, either iterative step-wise transport or small angle projection of charged particles is coupled to a Lorentz force solver. For the particle-pushing calculations in this work, PAFIN was equipped with a structure-preserving second-order integration scheme \cite{Hi2017} suitable for relativistic particle motion in electromagnetic fields \footnote{{Note: Equation (19) in ref.~\onlinecite{Hi2017} misses a factor of $c^2$ on the left hand side. Accordingly, we correct the last term in eq. (20) in ref.~\onlinecite{Hi2017} to $\| \vec{\beta} \cdot \vec{u} \|^2 \cdot c^{-2}$ instead of $\| \vec{\beta} \cdot \left(\vec{u}\right)^2 \|$.}}.\par


In this simulation, the parameters of the EM-mode are derived from the measured group velocity and the fit of the discharge pulse: we solve the 1D continuity equation assuming a constant group velocity $v_\mathrm{g}$,

\begin{equation} \label{eq:current}
I(x,t) = v_\mathrm{g} \cdot \lambda (x,t) .
\end{equation}

The discharge current $I (x,t)$ induces a B-field which co-propagates with the E-field. The temporal change of the fields are not explicitly taken into account for simplified simulations.\par

We perform dynamic simulations maintaining the fitted pulse shape, but re-normalizing it to different {total charges $Q_0$}. We compare the best fitting simulation results, in the bottom {row} of \cref{fig:timeresolved}, to their exact experimental counterparts in the top row. Note the perfect agreement of asymmetric features of focused particles in the coil center for early probing times. Some small deviation of simulated deflectograms and experimental results are visible at the vicinity of the $\Omega$-legs of the coil. We find \SI{Q_0=300}{nC} for the best overall agreement between experimental and synthetic deflectograms. The re-normalized peak (red curve) is compared to the original fit (blue curve) in \cref{fig:peak}. Note that an earlier analysis of the discharge stream around the omega shaped part of the target rod \cite{Eh:2016} pointed out that a charge density on the wire alone, creating electrostatic fields, does not {accurately} reproduce the experimental proton-deflections.\par

Simulations indicate that the streaming EM-pulses have amplitudes of tens of \SIU{GV/m} and tens of \SIU{T}. Note that \SI{10}{GV/m} and \SI{10}{T} in SI units is $\approx334000$ statV/cm and 100000 G in Gaussian units. In order to understand the influence of such low amplitude B-field of several \SIU{T} comparatively to electrostatic effects on the probing protons of several \SIU{MeV}, we track the protons crossing the loop along the symmetry axis. We analytically estimate an upper limit for the deceleration prior to the target by equating the potential of an uniformly charged ring and the kinetic energy:

\begin{equation}
\Delta v_{\parallel\mathrm{TNSA}} \leq 3 \sqrt{\lambda_\mathrm{  \SIU{nC/mm}}} \mathrm{[\SIU{\mu m/ps}]} .
\end{equation}

Protons of several \SIU{MeV} kinetic energy have velocities of several \SI{10}{\mu m/ps} -- with $\lambda \approx$ \SI{20}{nC/mm} it leads to $\Delta v_{\parallel\mathrm{TNSA}} \leq $ \SI{13}{\mu m/ps}. This relative change in velocity is non-negligible for the encountered field amplitudes. The electric component of the pulse is responsible for deceleration of particles prior to their {transit} through the coil - for one individual particle this ultimately results in a shorter length for focusing back to the axis by the effect of the magnetic lens corresponding to the loop current. 

After transiting through the loop, the E-field would lead to a longitudinal re-acceleration. In the case of a static E-field, this would keep the in- and out-going kinetic energy of particles approximately equal. Yet, as the charge density evolution is asymmetric, the decelerating and accelerating potential vary. In simulations, the difference of particle energies before and after passing the coil is of the order of hundreds of \SIU{keV}.\par

Dynamic simulations also reproduce better imprints of the deflections around straight sections of the conductor rod. Access to the full phase space of the probing particles gives further insight in the dynamic processes at the vicinity of the conductor. Before transiting by the wire, the particle decelerates in the direction parallel to the directrix of the hyperbolic particle orbit. This violates the first assumption made to derive the charge density. For probing protons at \SI{8.8}{MeV}, the change of kinetic energy reaches the order of \SI{1}{MeV}.\par

In summary, when the Weibull fit to the discharge profile {(blue curve in \cref{fig:peak})} is fed into PAFIN proton tracing simulations, the proton deflections are larger than in the experimental data. This suggests the EM field amplitude has been over-estimated by \cref{eq:lambda}. After re-normalization of the discharge profile {(red curve in \cref{fig:peak})}, however, we find that the PAFIN simulations agree very well with the experimental radiographs (see\cref{fig:timeresolved}).

\subsection{Ambiguity of Results}

Presuming the validity of the fit function \cref{eq:fit} with $\lambda(t) = f(v_\mathrm{g} \cdot t)$, the EMP peak is attained after a rise time of

\begin{equation}
\begin{aligned}
\tau_\text{rise} &= \sqrt[k]{\frac{k-1}{k}} \cdot \frac{\sigma}{v_\mathrm{g}} \overset{\text{exp}}{=} \SI{(11.3 \pm 1.2)}{ps} \gg \tau_\mathrm{L} \quad .
 \end{aligned}
\end{equation}

\noindent {As seen in \cref{fig:peak},} this rise time is about $22 \times$ the laser pulse duration. This result may be {a coupled effect of the target discharge in the {\it explosive} regime (as defined in ref.~\onlinecite{Po:2018}) on timescales longer than the laser drive and the} temporal resolution of charged particle beam deflectometry in the low-\SIU{MeV/u} projectile energy range. {We will refer to the charging dynamics later in this article.}\par

The discharge pulse travels with a velocity of about \SI{250}{\mu m / ps} which is approximately $5$ -- $25 \times$ faster than the probing protons at \SI{10}{\mu m/ps} for \SI{1}{MeV} kinetic energy and up to \SI{60}{\mu m/ps} for \SI{20}{MeV}. Dynamic simulations show that the propagating EM-fields affect protons passing as far as approximately \SI{250}{\mu m} distance from the wire. We see that the fastest protons are influenced by the discharge pulse for a duration corresponding to the rise time, as represented by the large uncertainty for the probing time in \cref{fig:peak}. The pulse may have a shorter rise time with steep spatial gradients of the potential yielding three dimensional deflections that are not covered by the analysis according to \cref{eq:lambda}.\par

This ambiguity on the leading edge of the discharge pulse claims for further investigations with a better temporal resolution for the peak, e.g. by using short laser-pulse probing for future experiments based on electro- and magneto-optic effects in thin film crystals \cite{Wi2002,Bi2017}. This is beyond the scope of the present paper.

\section{PIC Simulations}

{Our data analysis evidences} pulsed electric and magnetic field components streaming along the target rod. For a deeper understanding of the nature of the discharge pulse, we performed 2D PIC simulations of the laser-target interaction using the PICLS code \cite{Mi2013}. The simulations resolve the successive propagation of electromagnetic waves and accelerated particle species. First, we performed five times down-scaled {simulations (all sizes except the distance between the ends of the coil -- legs of the $\Omega$-shape)} to capture the whole target geometry {and} distinguish the various transient electromagnetic effects: the propagation of fast electrons, EMP {emission} and {guided} discharge pulse. In a second step, real-scale 2D PIC simulations are used to study the generation of hot electron current and return current as well as associated electromagnetic fields.\par

The down-scaled {simulations employ} a driver laser pulse at the intensity of \SI{10^{19}}{W/cm^2} comparable to the experiment but with \SI{0.8}{\mu m} wavelength and a pulse duration of \SI{1.33}{ps}. The pulse interacts with the target in normal incidence with a \SI{16}{\mu m} FWHM focal spot. Its electric field oscillates in the simulation plane. The spatial and temporal profiles are flat top with Gaussian edges. The target  plasma is composed of H ions and electrons with a initially uniform \SI{10}{n_c} density. The initial electronic and hydrogen temperatures are set to zero. The spatial step in both dimensions is \SI{40}{nm}. The time step is \SI{0.1}{fs}. The boundary conditions used are absorbing in both dimensions. Binary collisions and field ionization are not taken into account.\par

\begin{figure}[htb]
	\includegraphics[width=\columnwidth]{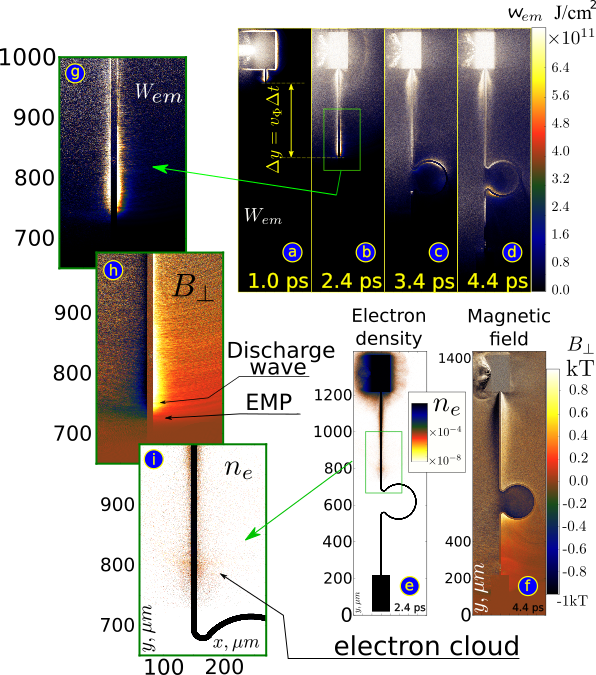}%
	\caption{Down-scaled 2D PIC simulations of the laser driven discharge on DCT show target-bound discharge pulse, EMP in free space and an accelerated electron cloud following the discharge peak along the target wire. Panels (a -- d,g) show the EM-energy-density in units of the 2D simulation, (f,h) the magnetic field strength perpendicular to the simulation plane and (e,i) the electron density. (g,h) highlight the spatially pulsed character of the discharge wave and (i) shows the accelerated electron cloud.
	\label{fig:discharge_PIC}}
\end{figure}

The resulting EM-energy-density $w_\text{em} = B_\perp^2/2\mu_0 + \epsilon_0 E_\parallel^2/2$ is given in \cref{fig:discharge_PIC} (a -- d,g) for different times. The driver {laser} pulse is injected at the left side of the simulation box. We see a discharge pulse bound to the target geometry at a high energy density, of several \SI{10^{11}}{J/cm^2}, and propagating at the velocity $0.964 c$, close to that of light. Its spatially pulsed character is highlighted {by} a zoom in \cref{fig:discharge_PIC} (g). The magnetic field amplitude perpendicular to the simulation plane is given for the same time (\SI{2.4}{ps}) in \cref{fig:discharge_PIC} (h). A spherical EMP in free space, that emanates from the interaction region with the velocity of light, is clearly distinguishable from the {guided} discharge pulse, which is slightly slower. For late times {of \SI{4.4}{ps}, \cref{fig:discharge_PIC} (f)}, we see strong magnetic fields in the vicinity of the {laser-plasma} interaction region. This indicates a return current building up.\par

The electron density \SI{2.4}{ps} after the interaction started is shown in \cref{fig:discharge_PIC} (e) and zoomed in \cref{fig:discharge_PIC} (i). Besides the plasma expansion in the hot interaction region, we identify a population of electrons that co-propagates with the discharge pulse.\par

The electric field streaming along the target has a monopole configuration in comparison to the fast oscillating EMP. The amplitude of the radial electric field at straight wire sections is \SI{100}{GV/m} in the simulation, scaling to \SI{20}{GV/m} in the experimental frame. The simulated amplitude of the magnetic field component around the target rod is \SI{500}{T}. This corresponds to \SI{100}{T} at the coil's center. Scaling to the experimental coil-size leads to \SI{\approx 20}{T} at the coil center. Both components agree in field strength with the values {heuristically deduced from experimental data} in the previous section. The PIC simulations confirm a very strong radial electric field, and a weak azimuthal magnetic field. Around straight wire sections the electric fled dominates the Lorenz force, in agreement with our initial assumption and supporting our evaluation of the guided EMP time evolution.


\begin{figure}[htb]
    \centering
    \includegraphics[width=\columnwidth]{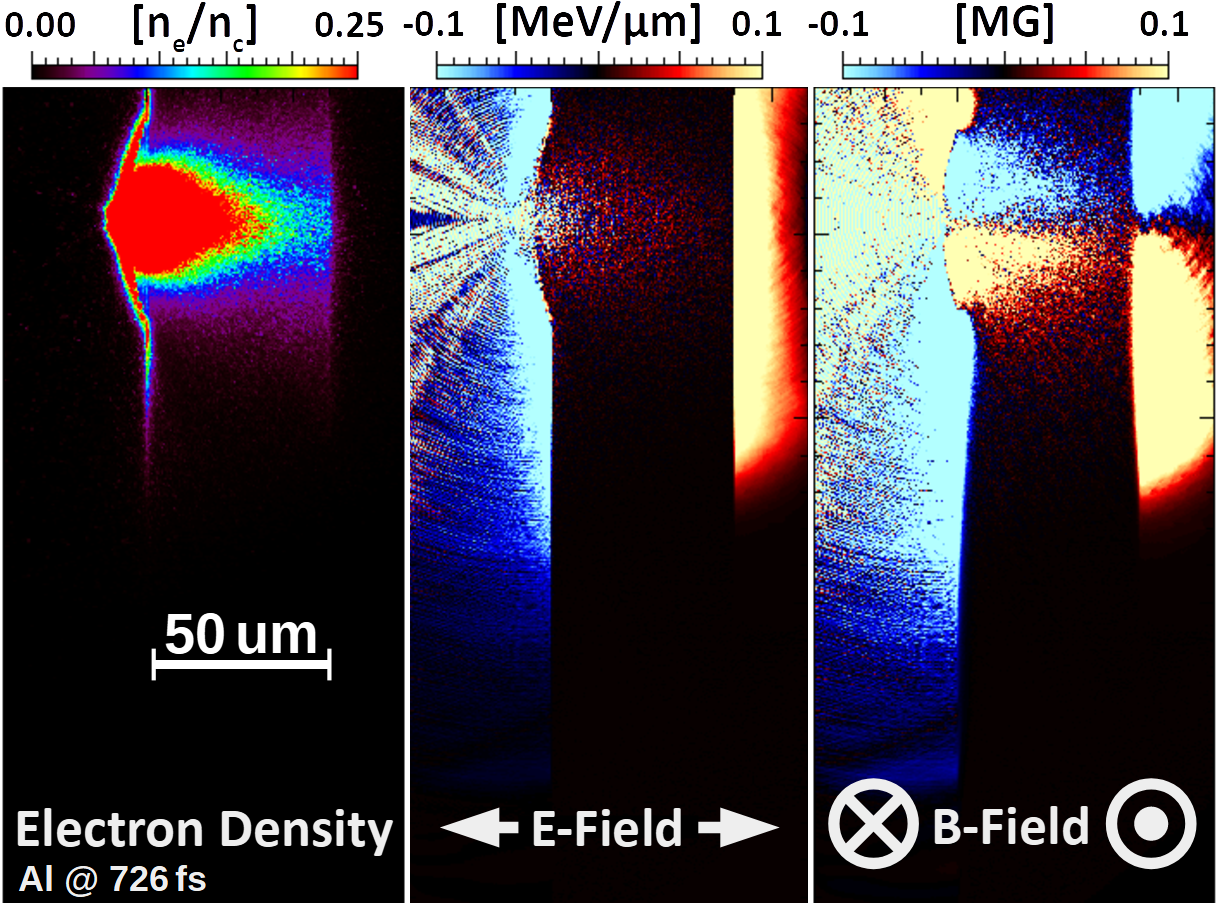}
    \caption{Real-scale 2D-PIC simulations showing the electron energy density, instantaneous $E_x$ electric field, with horizontal axis $x$ and instantaneous $B_z$ magnetic field, with $z$ the axis pointing out of the image plane, \SI{726}{fs} after the beginning of the simulation.}\label{fig:manupic}
\end{figure}

{More} detailed 2D PIC simulations are used to study the generation of hot electron current and return current as well as associated electromagnetic fields on real-scale \SI{50}{\mu m} {thick} foils. {Here,} we solely simulate the laser-target interaction region and a successive straight conductor section. The incident laser pulse with \SI{1}{\mu m} wavelength and a pulse duration of \SI{100}{fs} has a maximum intensity of \SI{10^{19}}{W/cm^2} within the \SI{3.5}{\mu m} FWHM of the focal spot. The pulse interacts with the target in normal incidence. Its electric field is in the simulation plane. The spatial and temporal profiles are truncated Gaussians. The target plasma is composed of Al ions and electrons with a \SI{700}{n_c} maximum density. A \SI{1}{\mu m } longitudinal scale-length exponential pre-plasma is present in front of the target with a Gaussian transverse profile and a total length of \SI{\Delta y=10}{\mu m}. The initial electronic and aluminum temperatures are set to zero. The spatial step in both dimensions is \SI{20}{nm} and there are $2$ Al ions and $26$ electrons per cell. The time step is \SI{0.066}{fs}. The boundary conditions used are absorbing in both dimensions. Field ionization using the ADK formula \cite{Perelomov1966,ADK:1986} is taken into account as well as impact ionization. Binary collisions are also taken into account.\par

Instantaneous magnetic and electric fields as well as the electron density are shown in \cref{fig:manupic}, \SI{726}{fs} after the beginning of the simulation. The laser pulse is injected at the left side of the simulation box. The maximum of the laser pulse enters the plasma after \SI{330}{fs}. We observe an azimuthal magnetic field $B_z$ of the order of \SI{100}{T} appearing inside the target as well as in the vicinity of the target rod. Even though there are electrons down-streaming the target from the laser-interaction surface, the orientation of the surface magnetic field is clearly indicating a positive charge density propagation. The $E_x$ electric field has a peak amplitude of several \SI{100}{GV/m}.\par

The simulation reveals different EM waves originating from the interaction region along both surfaces of the target: a spherical EMP in free space is visible on the front side, propagating with the velocity of light. A discharge pulse propagates along front and rear {target} surfaces. The different {progress} can be explained by the delayed build up of the potential at the target rear side. From the time evolution of the $B_z$ and $E_x$ fields along the target surface, the velocity of the downward propagating EM mode on the front surface is measured to be \SI{0.87}{c}. This is in good agreement with the experimental values, regarding both Cu and Al as similar perfect conductors. We measure the rising edge of the amplitude of the B-field in vicinity of the conductor and divide it by the group velocity of the pulse: the rise time {of \SI{(144 \pm 48)}{fs}} is of the order of the driver laser pulse duration {in the simulation}.\par


The dynamics of the {guided} EM pulse{, with} clear evidence of a spatial electromagnetic pulse with a mostly mono-mode transverse electric field structure, motivates analytical modelling {efforts} in order to conduct studies that do not require costly PIC simulations.

\section{Modelling of the Discharge Pulse}

We will compare the experimentally deduced target charging with numerical simulations in a first part and a second part will be devoted to explore the discharge wave dispersion for a better understanding of the group velocity difference to the speed of light.

\begin{figure}
	\includegraphics[width=.99\columnwidth]{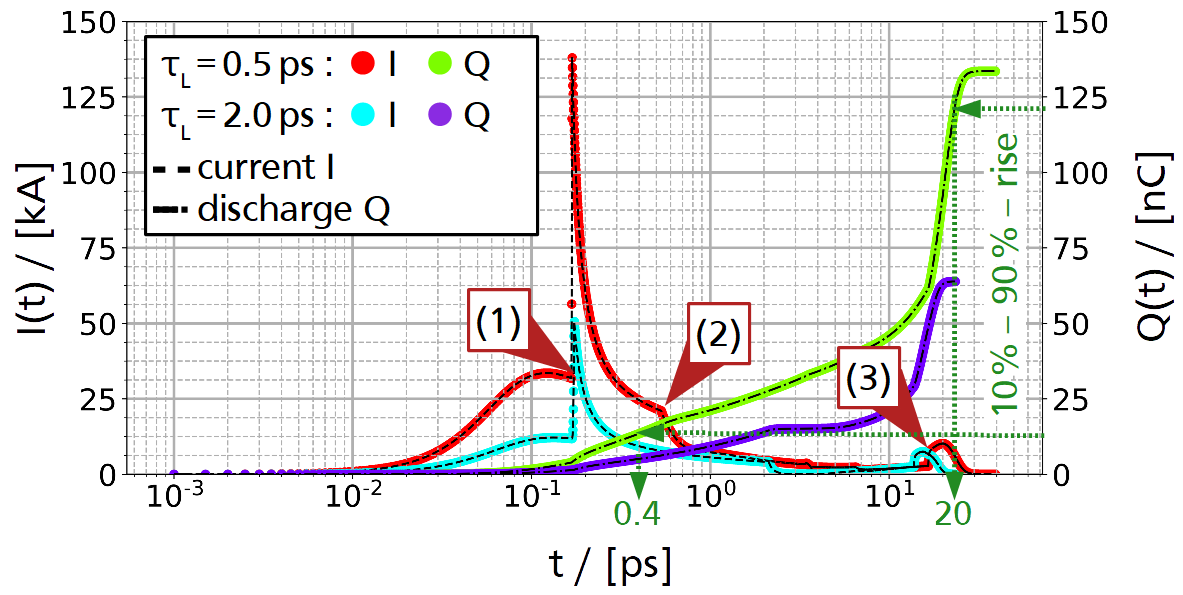}%
	\caption{Evolution of the target discharge current $I$ (red, cyan) and total charging $Q$ (green, violet) simulated with ChoCoLaT2 at $50\text{~\%}$ absorption efficiency but two distinct pulse durations of \SI{0.5}{ps} and \SI{2}{ps}. Numbered labels indicate (1) the occurrence of the target rear side contribution to the current due to electrons that crossed the target, (2) the end of the laser pulse and (3) the time when, due to collisional cooling and ejection of most energetic electrons, the average Debye length of the hot electron distribution in the target becomes larger than the target thickness. \label{fig:chocolatdischarge}}
\end{figure}

A first detailed attempt to model target charging in short laser pulse interactions \cite{Po:2015-3,Po:2018} allows to predict the expected discharge current due to laser-heated relativistic electrons. The numerical code ChoCoLaT2 simulates electron heating on a thin disk target and successive electron escape mitigated by the target potential, based on the driver laser parameters and the interaction geometry. It takes into account the collisions of electrons within cold solid density targets. The energy and time depending hot electron distribution function $f(E,t)$ evolves according to
\begin{align}
\partial_t f (E,t) &= \frac{h_\mathrm{L}(E) \Theta(\tau_\mathrm{L} - t)}{\tau_\mathrm{L}}  -  \frac{f (E,t)}{\tau_\mathrm{ee}(E)} - g(E,t)
\end{align}

\noindent where $h_\mathrm{L}(E)$ is a constant exponential source of hot electrons, $\Theta (t)$ the Heaviside function limiting electron heating to the laser duration, $\tau_\mathrm{ee}(E)$ the energy dependent cooling time and $g(E,t)$ the rate of electron ejection from the target. Source term and normalization evaluate with
\begin{align}
h_\mathrm{L}(E) &\overset{!}{=} \frac{N_0}{T_0} \exp{\left[ - E/T_0 \right]} \\
N_0 &\overset{!}{=} \int f (E,0) \text{~d}E \quad .
\end{align}

The initial hot electron temperature $T_0$ depends on laser wavelength and pulse intensity \cite{Fa1985,Be1997,Wi:1992}; and $N_0$ is re-normalized to the energy balance $N_0 T_0 = \eta E_\mathrm{L}$ between the total energy of hot electrons in the target and the absorbed laser energy.\par

We perform ChoCoLaT2 simulations using our interaction parameters and a laser absorption between \SI{30}{\%} and \SI{50}{\%}. Resulting $Q_0=$ \SI{(150 \pm 20)}{nC} are of the same order of magnitude as the experimental value. Taking into account that ChoCoLaT2 systematically underestimates target charge by a factor of 2 to 3 \cite{Po:2018}, we consider the agreement as fairly good.\par

\begin{figure*}[htb]
	\includegraphics[width=0.99\textwidth]{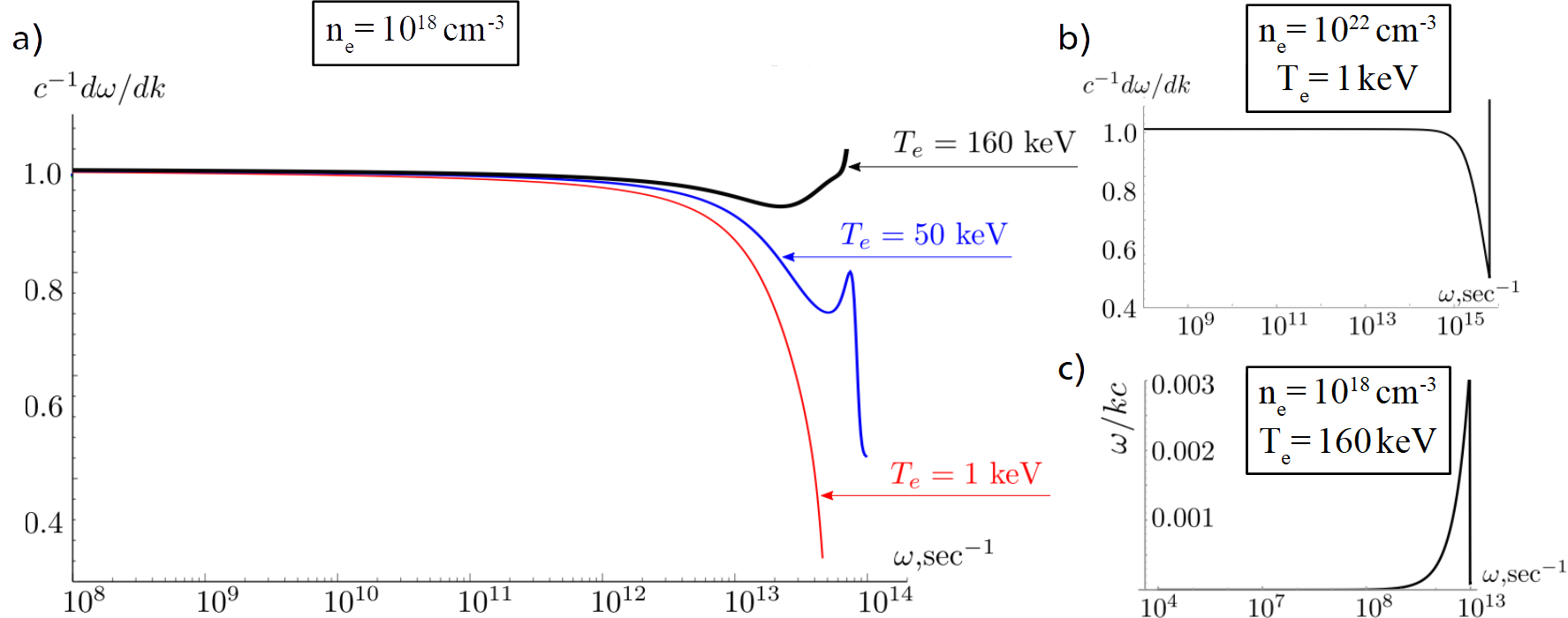}%
	\caption{(a) Frequency dispersion of the discharge pulse group velocity, calculated numerically from the dispersion relation \cref{eq:disp}, for \SI{n_e=10^{18}}{cm^{-3}} and \SI{T_e=1}{keV} (red, lower curve), \SI{T_e=50}{keV} (blue, middle curve), and \SI{T_e=160}{keV} (black, upper curve). (b) shows the numerical solution for the group velocity for \SI{n_e=10^{22}}{cm^{-3}} and \SI{T_e=1}{keV}. (c) shows the phase velocity for the lower branch of the dispersion relation \cref{eq:disp} for \SI{n_e=10^{18}}{cm^{-3}} and \SI{T_e=160}{keV}.
	\label{fig:velocity}}
\end{figure*}

A striking feature of the discharge wave propagation is its velocity different to the speed of light, with experimental data in \cref{tab:phasespeed}. To understand this interesting phenomenon, consider the wire as a plasma cylinder with radius $a$, temperature $T_e$ and electron density $n_e$. The electromagnetic wave propagation is considered using Maxwell equations in the cylindrical coordinate system $(r,\theta,z)$ with unit vectors $\vec e_r,\vec e_\theta,\vec e_z$, both inside and outside the plasma cylinder

\begin{equation}
\begin{cases}
\left[\partial_z E_r(r,z,t) - {\partial_r E_z(r,z,t)} \right] \vec{e}_{\theta}= -\frac{1}{c}{\partial_t \vec{B}(r,z,t)} , \\
\frac{1}{r}{ \partial_r \left[ r  B_{\theta}(r,z,t) \right]} \vec{e}_{z} - {\partial_z B_{\theta}(r,z,t)} \vec{e}_{r}= \frac{1}{c}{\partial_t \vec{D}(r,z,t)} ,\\
\frac{1}{r}{ \partial_r \left[ r  D_{r}(r,z,t) \right]} + {\partial_z D_{z}(r,z,t)}= 0.
\end{cases}
\label{maxwell}
\end{equation}
Plasma properties are defined by the dielectric tensor, which non-zero components, in the simple case of Maxwellian collisionless plasma, read \cite{LL10engl}
\begin{equation}
\varepsilon_{rr}(\omega,k)=
1-\frac{\omega_e^2}{\omega^2}F\left( \frac{\omega}{\sqrt{2}kv_T} \right),
\label{eq:epsrr}
\end{equation}
\begin{equation}
\varepsilon_{zz}(\omega,k)=
1+\frac{\omega_{e}^2}{(kv_T)^2}\left[1+F\left( \frac{\omega}{\sqrt{2}kv_T} \right) \right],
\label{eq:epszz}
\end{equation}
where
\begin{equation}
F(x)=\frac{x}{\sqrt{\pi}}\lim\limits_{\delta\to0}\int\limits_{-\infty}^{\infty}\frac{e^{-z^2}}{z-x-i \delta}dz,
\label{eq:epsF}
\end{equation}
and $\omega_e=\sqrt{4\pi n_e e^2/m_e}$ is the electron plasma frequency, $v_T=\sqrt{T_e/m_e}$ is the thermal electron velocity, $m_e$ is the electron mass.\par

To obtain the dispersion relation, the field components are transformed into their Fourier transform components,
\begin{multline}
\{E,B,D\}_i(r,z,t)= \\ \int \{E,B,D\}_i(r,k,\omega) \cdot \mathrm{e}^{-i (\omega t + k z)} dk d\omega ,
\end{multline}
and substituted to \cref{maxwell}, which provides a set of the second-order differential equations for cylindrical functions. The solutions should be finite at $r\to0$ and $r\to\infty$, and also they must be joined at the edge of the plasma cylinder $r=a$.
The consistency of all these conditions provides the dispersion relation
\begin{equation}
\frac{K_0(\alpha k a)}{K_1(\alpha k a)}=-\frac{\left(\frac{1}{\varepsilon_{rr}(\omega,k)}-\frac{\omega^2}{k^2c^2}\right)}{\alpha \beta}\frac{I_0(\beta k a)}{I_1(\beta k a)},
\label{eq:disp}
\end{equation} 
where $\alpha\equiv \pm\sqrt{1-\frac{\omega^2}{(kc)^2}}$, $\beta\equiv\pm\sqrt{\frac{\varepsilon_{zz}(\omega,k)}{\varepsilon_{rr}(\omega,k)}-\varepsilon_{zz}\frac{\omega^2}{(kc)^2}}$, $I_i(x)$ and $K_i(x)$ are $i$-th order modified Bessel functions of the first and the second kind respectively.\par

From the PIC simulations, we may conclude that the {considered cylindrical} plasma is hot and {its initially sharp edges} diffuse on the scale of the Debye length. It is possible to make only qualitative conclusions from the model dispersion \cref{eq:disp}, using the effective electron plasma density of the hot layer around the original solid-density cold wire. According to the PIC simulations in \cref{fig:discharge_PIC} the effective electron density is of the order of \SI{n_e\sim10^{18}}{cm^{-3}}. Assuming the main frequency of the discharge wave to be of the order of the inverse laser pulse duration $\omega\sim 2\pi/\tau_L$, \SI{\tau_L=0.5}{ps}, that is \SI{\omega\approx 1.2\times10^{13}}{s^{-1}}, we find that the plasma frequency for this effective electron density is somewhat higher \SI{\omega_e\approx 5.6\times10^{13}}{s^{-1}}. The difference between these two frequencies is an important parameter, which can explain an observed group velocity considerably lower than the light velocity. We can see in \cref{fig:velocity} (a), for \SI{n_e=10^{18}}{cm^{-3}}, that the group velocity is about 10\% less than the light velocity in the domain $\omega\sim \SI{10^{13}}{s^{-1}}$. This decrease is due to the decrease of the plasma frequency for low-density plasma close to the critical frequency for the propagating pulse.\par

Another important parameter in the model is the plasma temperature. It actually defines the rate of collisionless Landau damping, which is growing up with plasma temperature as more resonant particles are present in the system. In \cref{fig:velocity} (a) three curves show the group velocity, numerically calculated from \cref{eq:disp} for three plasma temperatures. For a relatively low temperature \SI{T_e=1}{keV} (red curve), the resonance at plasma frequency $\omega=\omega_e$ is very sharp, and the group velocity goes down quite deeply, to $\omega/(kc)\sim 0.6$. Increasing of the temperature results in smoothing of this discontinuity, {as illustrated by the blue} curve for \SI{T_e=50}{keV}. For the parameters of the actual experiment, however, the scale of the hot electron energies in PIC simulations is about hundred \SIU{keV}. For this situation, the curve becomes very flat, decreasing at it most to $\omega/(kc)\sim0.8 - 0.9$ (black curve), consistent with the experimental observation within error bars. Nevertheless, PIC simulations may overestimate the temperature, and lower temperatures can also explain the drop to values below \SI{80}{\%} that were experimentally observed.\par

Note that comparing \cref{fig:velocity} (a) and (b), the lower the electron density is, the lower is the group velocity for a given frequency. This result may have already been obtained, if one considers just the propagation of the wave along a cold copper wire \cite{batygin}. Thus, the highly nonlinear behavior of the group velocity is defined by the two main parameters; effective plasma density and temperature, both sensitive to the irradiation conditions. This may give rise to the variation of the experimentally observed values of the phase velocity, and motivates further experimental studies.\par

Landau damping is the only absorption mechanism, which makes the wave to be not purely transversal. This effect may contribute to effective electron acceleration along the wire \cite{McKenna-prl2007, Kuratov2018}, as seen in \cref{fig:discharge_PIC} (i).

\begin{figure*}[htb]
    \centering
    \includegraphics[width=\textwidth]{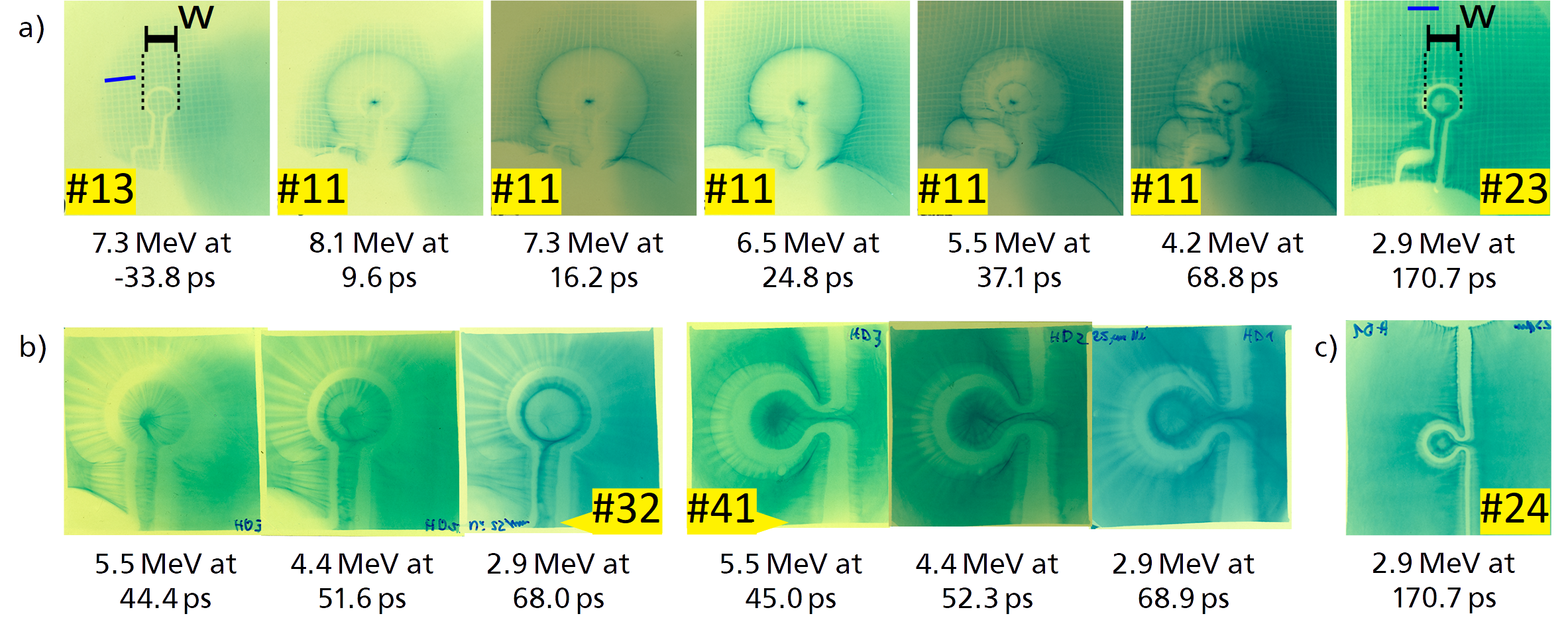}
    \caption{Typical evolution of deflections imaged on RCF for DPCT and DCT geometries. For DPCT in (a), the comparison of an early imprint (for shot \#13) and a late imprint (for shot \#23) with the magnification corrected scale-bar (w) as reference, we see that the coil rod increased by the order of two conductor widths, that is up to \SI{100}{\mu m}. (b) For the exponentially decaying discharge pulse amplitude after passage of the peak at \SI{\approx 12}{ps}, the intense imprint in the coil center issued from the dense beam prgressively faints, while a ring shaped caustic appears, concentric with the coil. Both shots \#32 and \#41 in (b) were performed with increased magnifications by approaching the TNSA source to \SI{1.99}{mm} and \SI{2.01}{mm} from the coil plane respectively. DPCT deflectograms (shot \#32) agree with the ones obtained for DCT geometry (shot \#41), (c) also for latest probing times (shot \#24). The center void within the ring shaped caustic lasts for our latest observations. Note the coil shadow in each imprint, the coil diameter is \SI{500}{\mu m} in the coil plane, blue bars in (a) indicate the distance of \SI{6.5}{mm} in the detector plane. The timing reference is the driver laser impact, proton energies are given, and probing times account for the time of flight from source foil to coil plane.}\label{fig:evolution}
\end{figure*}

\section{Dynamics after the Discharge Pulse}

Analysis of the proton-imprints in \cref{fig:evolution} and \cref{fig:currentevolution_DCT} allows to detail the evolution of the electromagnetic effects for later times. Following the full evolution for DPCT geometry in \cref{fig:evolution} (a), the long tail of the discharge pulse continues to weakly squeeze the charged particle beam in proximity of the loop, but two striking changes arise: the appearance of a 'sun-rayed' pattern of caustics in vicinity of the coil and a doughnut shaped caustic inside the coil. Besides these two characteristic caustics, we diagnose the rise of the return current by {the proton deflections around the coil}, when probing perpendicularly to the coil axis (\cref{fig:currentevolution_DCT}).

\subsection{Characteristic Caustics}

A `sun-rayed' pattern of proton density minima is visible on RCF imprints. It appears inside and around the coil and the stripes are perpendicular to the conductor surface. The perpendicularity is especially pronounced in the $\Omega$-leg part for shot \#32 with a DPCT, as shown in \cref{fig:evolution} (b). Such caustics are observed in all shots, \SI{(33 \pm 11)}{ps} after the passage of the pulse peak on the coil. The deflection pattern remains stationary, caustics change contrast but not their location with respect to the conductor. The hydrodynamics of a wire plasma is too slow at the estimated heating rate to form a modulated plasma density at the observed distance around the wire. The observed ray-like structure is therefore probably defined by a modulation of the potential on the conductor or in direct vicinity of the target.\par

Variations in the potential might be caused by the rising return current, {as studied in} \cite{Po:2015-1}, Appendix D. That paper describes such fluctuations, without taking into account the retarded character of the evolving fields. Assuming a constant propagation speed of the pulse with $(0.82 \pm 0.06) \cdot c$ and a spread of retarded feedback with $c$, we obtain an estimated time-of-travel from interaction region to grounding and back to the coil of \SI{(30.8 \pm 4)}{ps}. Considering our target mounting with a conductive glue drop of \SI{\approx 1}{mm} diameter that holds the target on the grounding needle, time-of-travel and development of the caustic pattern overlap in the range of their uncertainty.\par

Another possible explanation for the ray pattern would be a modulation of the discharge wave itself. According to the model presented in the previous section for the discharge wave dispersion, the phase velocity of a short scale modulation of a Sommerfeld-like propagating wave appears to be very low. Accordingly, the ray-pattern would be almost constant during the observation time. In this case, consider a low-velocity branch of the solutions of \cref{eq:disp}. In the limits $\omega\to0$, $\omega/(kc)\to0$, the dispersion relation gives to first order 
\begin{equation}
(k a)^2\left( \gamma+\ln\frac{2}{k a} \right)\pm2\left(\frac{k v_T}{\omega_e}\right)^2\approx0,
\label{eq:lowlimit}
\end{equation}

\noindent where $\gamma\approx 0.577$ is the Euler-Mascheroni constant. The constant value of the wave number $k\approx 1.12 a^{-1}$, defined from \cref{eq:lowlimit}, of the order of the inverse cylinder radius, corresponds to the low-velocity branch of the discharge wave. It may be excited if a seed perturbation, e.g. from the surface wire structure, is applied along the plasma cylinder. Comparing the spatial \SI{\sim 50}{\mu m} and temporal \SI{\sim 30}{ps} scales of the fine ray-like structure in \cref{fig:evolution}, we see, that this solution provides its qualitative description.\par

During the emergence of the ray-pattern, \cref{fig:evolution} (b), the contrast of the beam on the coil axis becomes weaker, then progressively a ring-shaped sharp cusp appears with increasing contrast. The ring is concentric with the coil and clings close to the conductor on the inner side. For even later probing, no more protons reach the RCF on the coil axis and a clear void forms (see \cref{fig:evolution} (c)). Void and ring are visible for the latest observation times, at \SI{\approx 171}{ps} after the laser-interaction. The ray pattern is barely visible already after \SI{\approx 70}{ps}. The ring stays very clear. The evolution from focus point towards a strong ring-like deflection arises independently of the target geometry. Possibly, electrons coming from the laser interaction region get trapped in the vicinity of the coil and perturb proton deflectometry results.\par

After the passage of the full discharge wave, the proton image of the target appears nearly un-altered. Only for some shots, target rods are up to twice as large compared to images of the yet undriven target. The initially straight rod shows small surface modulations. Eventually ohmic heating \cite{Ti:2017} led to a slight expansion of the target wire with a velocity of  \SI{\approx (0.5 \pm 0.3)}{\mu m/ps}, or we see deflections imposed by a slightly charged target with hundreds of \SIU{pC/mm}. Comparing an early imprint for shot \#13 and a late imprint for shot \#23 in \cref{fig:evolution} (a) with the magnification corrected scale-bar (w), we see that the coil diameter {slightly} increased by the order of two conductor widths, that is \SI{100}{\mu m}.

\subsection{Magnetic Field Signature}

\begin{figure*}[htb]
    \centering
    \includegraphics[width=\textwidth]{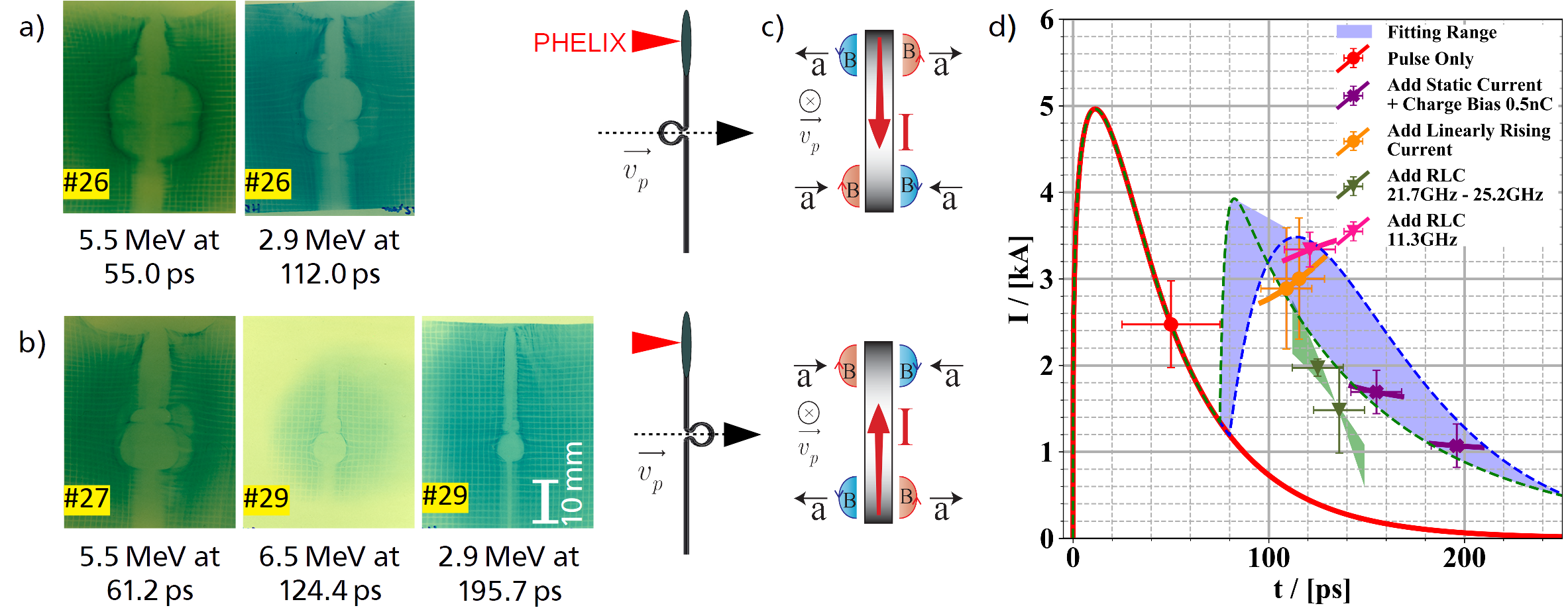}
    \caption{Current evolution based on RCF deflections for flat $\Omega$-Coil Target geometry. Side-on deflectometry, where coil axis and probing axis are perpendicular, with (a) $\Omega$-Coils and (b) $\Omega$-legs facing the proton source side, yields a characteristic bulb shape due to the (c) differing Lorentz force in top- and bottom-sections of the coil (the direction of current is indicated for the side facing the source). Through (a--c) top and bottom rows show results of one versus the other orientation. The timing reference is the driver laser impact, proton energies are given. (d) Return current amplitudes are obtained by fitting data to dynamic synthetic deflectometry simulations. Early deflectographs can be reproduced with the discharge pulse alone (red curve and point), then a mismatch arises that can be resolved assuming static- (purple), linear- (orange) and sinusoidal currents (green and pink). Data points correspond to the best fitting amplitude within a temporal evolution plotted by a line or interval in the same color. The later temporal evolution is plotted for the time of the passage of the probing proton in vicinity of the coil. Fits to model \cref{eq:modelPDC} yield a range of results plotted as 'fit range', probably based on shot-to-shot variations.}\label{fig:currentevolution_DCT}
\end{figure*}



Where early probing times, \SI{<60}{ps}, indicate B-fields induced by the displacement current of the discharge pulse, late probing times unravel a superposed charge-neutral return current. The magnetic field signature is more clearly evidenced in shots where the probe beam symmetry axis is set perpendicular to the axis of the omega-shaped coil (see \cref{fig:currentevolution_DCT}, where (a) and (b) correspond to DCT with the $\Omega$-Coil or alternatively the $\Omega$-legs facing the proton beam source, respectively). In such configurations, at earlier probing times, E-fields deflect symmetrically all protons passing above or below the coil's symmetry axis. With probing times greater than \SI{60}{ps}, we start observing an asymmetry comparing both orientations. Such polarity in the deflection is a signature of B-fields, see \cref{fig:currentevolution_DCT} (c). Bulb shaped caustics of this size are a widely observed phenomenon for ns-scale laser drive \cite{Sa:2015,Br2020,Pe2020}, and with the discussed target dimensions a clear indication of strong return currents.\par

{Note, that shot \#27 in \cref{fig:currentevolution_DCT} (b) witnesses a sudden change from symmetric to asymmetric deflections for probing at $t=$ \SI{61.2}{ps}, a weak symmetric caustic is superposed by a strong asymmetric caustic. The field configuration may have changed rapidly during the passage of the probing particles.}\par

The rise of a return current during the discharge wave decay is consistent with the target geometry. The discharge pulse reaches the grounding glue drop at $t=$ \SI{24 \pm 4}{ps}, then spreads over a large area and drops in charge density, accordingly. Electrons from the target holder are eventually causing a return current to rise. This return current could reach the coil already at $t=$ \SI{30.8 \pm 4}{ps}.\par

The bulb imprint on deflectograms does not change its polarity between passage of the discharge pulse and late times. This observation suggests there is a charge-neutral return current coming from the ground rather than a significant reflected charge pulse. There is no change in polarity during the full observation time of \SI{\approx 200}{ps}.\par

Dynamic simulations with PAFIN reproduce the deflections for early times employing a linearly rising return current with $\mathrm{d}_t I = \SI{20}{kA/ns}$. The simulations take into account the wave character of the current in the vicinity of the coil. Bulb sizes indicate currents of several \SIU{kA}, shown in \cref{fig:currentevolution_DCT}(d). Deflectograms of latest probing times at \SI{(150 - 200)}{ps} can be well reproduced with the B-field of a quasi-static current in the target superposed to the E-field of a charge density bias on the wire that has the order of \SI{0.5}{nC/mm}. This order of magnitude for a residual charge density is consistent with the deflections observed around straight sections of the target rod, discussed previously. Integrated over the full target surface, the total residual target charging is estimated to be \SI{\approx 20}{nC}.\par

Linearly evolving or quasi-static return currents do not yield good agreement between experimental data and synthetic deflectograms for probing times of \SI{100}{ps} to \SI{140}{ps}, suggesting instead transient current dynamics. The return current $I_\mathrm{rc}$ forms due to the residual potential on the target. Assuming a lumped element RLC-circuit with the same current in the full wire part of the target, the governing equation is
\begin{equation} \label{eq:modelRLC}
 \partial_t^2 I_\mathrm{rc}(t) + 2\delta_\mathrm{RL} \partial_t I_\mathrm{rc}(t) + \omega_\mathrm{LC}^2 I_\mathrm{rc}(t) = 0\ .
\end{equation}

\noindent with $\omega_\mathrm{LC} = 1/\sqrt{LC}$ and $\delta_\mathrm{RL} = R/(2L)$.

A flat copper DCT is modeled according to its geometry with a inductance of \SI{L=7.25}{nH}, a resistivity of \SI{\eta=16.8}{n \Omega m} and a capacity of \SI{C_1=222.5}{fF}. In the extreme case of an oxidized target consisting of CuO, assuming the same resistivity but at higher permitivity of $\epsilon_\mathrm{rm} = 18.1$, \SI{C_2=1.28}{pF}.\par

Taking into account the skin depth \cite{Jo1963}, one obtains ${\omega_\mathrm{RLC} = \SI{25.15}{GHz}}$ and ${\delta_\mathrm{RL} = \SI{34.43}{MHz}}$. For a fully oxided CuO target one obtains instead ${\omega_\mathrm{RLC} = \SI{5.91}{GHz}}$ and ${\delta_\mathrm{RL} = \SI{16.96}{MHz}}$. For both cases, these numbers represent underdamped oscillations with ${\omega_\mathrm{LC} >> \delta_\mathrm{RL}}$. The damping factor indicates a \SIU{ns}-scale current dynamics. Therefore we {would} expect an oscillation with frequency $\omega_\mathrm{RLC}$ with periods ranging from \SI{\approx 250}{ps} to \SI{\approx 1}{ns} depending on the degree of oxidation. PAFIN simulations in this frequency range reproduce particular proton imprints in only some of all the shots, see \cref{fig:currentevolution_DCT}(d), but there is no possible fitting for all data points. 

Instead, we observe a pulsed character of the current, which suggest the overdamped regime of the RLC system, ${\omega_\mathrm{LC} << \delta_\mathrm{RL}}$. Thus, a more accurate modeling is undertaken with the solution of a Pulse Discharge Current (PDC), with

\begin{equation} \label{eq:modelPDC}
 \begin{aligned}
 I_\mathrm{PDC}(t) &= \frac{Q_0}{\alpha^{-1} - \beta^{-1}} \cdot \left( \exp{\left[ -\alpha t \right]} - \exp{\left[ -\beta t \right]} \right) \\
 \alpha &= \delta_\mathrm{RL} \pm \sqrt{ \delta_\mathrm{RL}^2 -  \omega_\mathrm{LC}^2 } \\
 \beta &= 2  \delta_\mathrm{RL}  - \alpha
  \end{aligned}
\end{equation}

The PDC model fits to the data in a wide range of parameters, as illustrated in \cref{fig:currentevolution_DCT}(d). The range where valid fit functions can be produced is indicated as the blue shaded area. The parameter $\alpha$ ranges from \SI{\approx 25}{GHz} to \SI{\approx 420}{GHz} and $\beta$ then results inversely proportional with values from \SI{25}{GHz} to \SI{10}{GHz}. As the shape of the target visibly does not change with probing time, one may assume a constant inductance. With $L$, $\alpha$ and $\beta$ set, the PDC model allows to calculate ranges for $R$ and $C$, with

\begin{equation} \label{eq:modelPDC_solve}
 \begin{aligned}
 R &= L \cdot \left( \alpha + \beta \right) \\
 C &= \left( L \alpha \beta \right) ^{-1} .
  \end{aligned}
\end{equation}

The resistance ranges from \SI{363}{\Omega} to \SI{3118}{\Omega} and the capacitance from \SI{220.7}{fF} to \SI{32.84}{fF}, respectively. The latter suits well the capacity of a pure copper target, with a better agreement for the higher end of the interval. The corresponding resistance value of \SI{363}{\Omega} indicates a resistance two orders of magnitude above the case of cold copper with \SI{R \approx 5}{\Omega} for tens of \SIU{GHz}.\par

\begin{figure}[htb]
    \centering
    \includegraphics[width=\columnwidth]{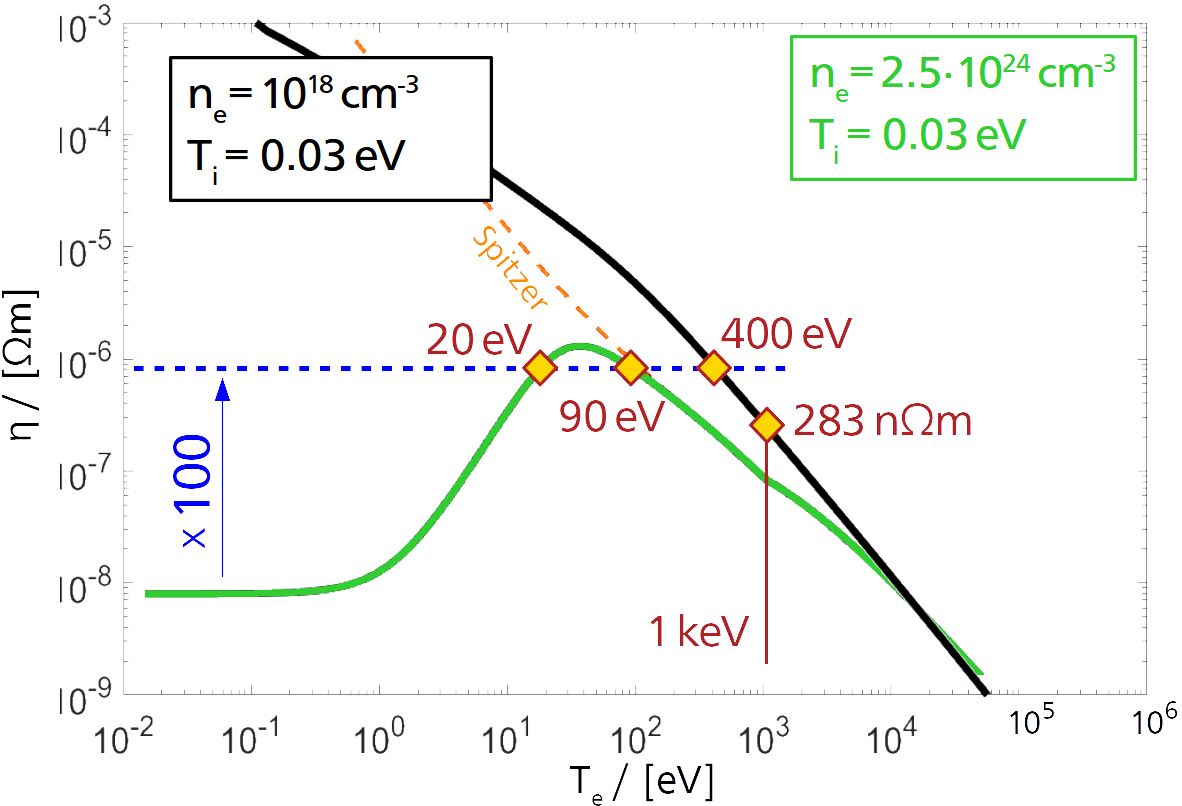}
    \caption{Resistivity calculated for Copper with cold background ions at different electronic densities. The green solid line illustrates the resistivity of solid density Copper, the dashed line indicates the Spitzer resistivity for this case. The black solid line depicts the Eidmann-Chimier resistivity for Copper at a density value discussed to explain the discharge wave dispersion.}\label{fig:resistivity_ch2008}
\end{figure}

This increase of resistivity can be a further indication that the target is heated. A large increase in resistivity $\eta = m_\mathrm{e} \nu_\mathrm{e} (T_\mathrm{e}) / q_\mathrm{e}^2 n_\mathrm{e}$ can be reasoned by the temperature depended electron collision frequency \cite{Ch2007} and mutually low electron densities. For an electron density of $n_\mathrm{e} = $ \SI{10^{18}}{cm^{-3}} as seen in PIC simulations, {and electron temperatures of \SI{1}{keV} reasonable to explain the group velocity of the pulse (see \cref{fig:velocity}),} the resistivity does increase to values that explain the large resistance, see \cref{fig:resistivity_ch2008}. An increase by the exact factor of $100$ is calculated for a slightly lower temperature of \SI{400}{eV}. Note that higher electron densities require lower temperatures to reproduce observations with our modelling for both the group velocity and the resistance.\par

Note further that surface plasma may change both the inductance and capacitance of the conductor. This underlines the importance of further studies aiming at experimental determination of the physical properties of the conductor.

\section{Conclusion}

{Our experiment has revealed} pulsed \SIU{kA}-currents on the timescale of tens of \SIU{ps} dispersing on laser-driven discharge targets. The velocity and dispersion of prompt discharge pulses indicate that a hot surface plasma forms on the wire section that connects the target to ground.\par

We see that the temperature and electron density of the surface plasma are promising control parameters of the discharge pulse dispersion. The dispersion relation is responsible for a group velocity different from that of light. Solutions on the low branch of the dispersion relation agree with modulations of the target potential in their spatial dimensions and temporal growth rate. Even if for this experiment, the seed of the potential modulation is not being controlled, their imprint on the \SIU{MeV} protons is clearly visible. Further studies are necessary regarding the origin of the surface plasma, the discharge pulse dispersion relation and controlled seeding of potential modulations.\par

The laser-driven EM discharge pulse with amplitudes of tens of \SIU{nC/mm} and several \SIU{kA} precedes the return current in form of a pulsed discharge current with several \SIU{kA}. We, for the first time, experimentally separate both currents with a well defined $\Omega$-loop shaped feature in the target rod. PIC simulations allow to distinguish EMP, fast electrons and a target-surface discharge wave propagation.\par

Building on this work, we see that relatively simple, flat metallic targets can be used for the chromatic lensing of charged particle beams. Using a dual laser set-up, energy-selection of the focused particles is possible by tuning the delay between the laser pulse driving the coil and the one generating the proton beam.\par

In the literature, comparable laser driven platforms are reported for the generation of pulsed magnetic fields \cite{Zh2018}, and the tailoring of laser-driven particle beams \cite{Ka:2016,Ah:2021}, but with no separation {or identification} of both transient currents. Note, that a parametric study of the discharge pulse parameters has been carried out recently \cite{Ak2019}, investigating charge density maximum and integral charge {as a} function of laser pulse duration, pulse energy and pulse intensity. Higher magnetic fields may be expected in a similar, but a more compact setup, where the loop itself is irradiated at one of its ends and the discharge current is closed due to the expanding plasma \cite{Ko2022}. A partial characterization of the pulsed discharge current has been carried out in ref.~\onlinecite{Wa2014}, also demonstrating neutral \SIU{kA} currents. A detailed exploration of the discharge pulses discussed in this paper is important for a range of applications in laser physics and laser-driven charged particle beam acceleration, particularly for medical applications, the heating of material samples to warm dense matter conditions using ion beams and the fast ignition approach to fusion.

\begin{acknowledgments}
We want to thank our funding projects POPRA Proj. 29910, IdEx U-BOR and CRA-ARIEL. PhK acknowledges support from the project \# FSWU-2020-0035  Ministry of Science and Higher Education of the Russian Federation. This work was granted access to the HPC resources of CINES under the allocations 2016-056129 and 2017-056129 made by GENCI (Grand Equipement National de Calcul Intensif). The experimental work has been partially carried out within the framework of the EUROfusion Consortium and has received funding from the Euratom research and training program 2014-2018 and 2019-2020 under grant agreement No 633053. The views and opinions expressed herein do not necessarily reflect those of the European Commission.
\end{acknowledgments}


\bibliography{manuscript}

\end{document}